# Virtual Critical Coupling

Younes Ra'di[1*], Alex Krasnok[1*], and Andrea Alú[1,2,3]

[1]*Photonics Initiative, Advanced Science Research Center, City University of New York, NY 10031, USA*

[2]*Physics Program, Graduate Center, City University of New York, NY 10016, USA*

[3]*Department of Electrical Engineering, City College of The City University of New York, NY 10031, USA*

To whom correspondence should be addressed: aalu@gc.cuny.edu

*These authors contributed equally

**Abstract**

Electromagnetic resonators are a versatile platform to harvest, filter and trap electromagnetic energy, at the basis of many applications from microwaves to optics. Resonators with a large intrinsic quality factor ($Q$) are highly desirable, since they can store a large amount of energy, leading to sharp filtering and low loss. Exciting high-$Q$ cavities with monochromatic signals, however, suffers from poor excitation efficiency, i.e., most of the impinging energy is lost in the form of reflection, since high-$Q$ resonators are weakly coupled to external radiation. Although critical coupling eliminates reflections in steady-state by matching the intrinsic and coupling decay rates, this approach requires the introduction of loss in the resonator, causing dissipation and lowering the overall $Q$-factor. Here, we extend the notion of critical coupling to high-$Q$ lossless resonators based on tailoring the temporal profile of the excitation wave. Utilizing coupled-mode theory, we demonstrate an effect analogous to critical coupling by mimicking loss with non-monochromatic excitations at complex frequencies. Remarkably, we show that this approach enables unitary excitation efficiency in open systems, even in the limit of extreme quality factors in the regime of quasi-bound states in the continuum.

*Introduction* – Resonant cavities are key components in various technologies and applications, including lasing, cavity electrodynamics, harmonic generation and sensing [1,2]. The ability of a cavity to store energy is usually described by its intrinsic quality factor [3], defined as $Q_\mathrm{i} = \omega_0 / 2\gamma_\mathrm{i} = \omega_0 / [2(\gamma_\mathrm{r} + \gamma_\mathrm{nr})]$, where $\omega_0$ is the resonance frequency and $\gamma_\mathrm{i}$ is the decay rate, which includes both radiative ($\gamma_r$) and nonradiative ($\gamma_{nr}$, i.e., dissipation or decoherence) portions, $\gamma_i = \gamma_r + \gamma_{nr}$. Consequently, the intrinsic quality factor can be expressed as $Q_i^{-1} = Q_r^{-1} + Q_{nr}^{-1}$ [4], where $Q_r$ and $Q_{nr}$ are responsible for leakage and dissipation, respectively. In addition to large storing capacity, high-$Q$ systems are highly desirable since they can provide enhanced wave-matter interactions, of importance for various technologies, e.g., lasers with lower thresholds and sensors with higher sensitivity. The widespread demand for high-$Q$ resonators has inspired significant research and various proposed layouts, such as microdisks [5–9], microspheres [10], Bragg reflector microcavities [11], and photonic crystals [12–14] with high $Q$ factors ($\sim 10^3 - 10^6$). In the same context, a novel class of optical cavities supporting bound states in the continuum (BICs) and consequently enabling unbounded intrinsic quality factors has been introduced [15,16].

In many applications, the cavity energy does not need to be converted, but only stored or delayed for a given amount of time. However, there are fundamental limits on the efficiency with which the energy can be delivered into them. As an example, let us consider an ideal cavity with no dissipation and radiative loss (i.e., $\gamma_r, \gamma_{nr} = 0$) connected to one scattering channel and excited by a monochromatic wave *via* this channel. In this scenario, due to the mismatch between the intrinsic decay rate ($\gamma_i = 0$) of the cavity and the coupling rate of the excitation channel ($\gamma_w$), i.e., $\gamma_w > 0$, the high-$Q$ resonator is not impedance-matched to the excitation channel. The quality factor of the system in this case reads $Q_a = \omega_0 / [2(\gamma_i + \gamma_w)] = \omega_0 / 2\gamma_w$. Throughout the transient regime, a significant portion of the external excitation is reflected at the cavity input, which results in an inefficient excitation process. After the cavity fills up, i.e., at the end of transient regime, all the incoming wave will be fully reflected, making the steady state an undesirable regime to store energy.

To improve the energy storage efficiency throughout the transient regime, one may think of using the concept of *critical coupling* (or quality factor matching). To this end, the intrinsic decay rate needs to be adjusted to match the coupling rate of the excitation channel, which, for a single port resonator, simply means adding proper amount of dissipation loss into the cavity. The quality factor of the system in the critical coupling regime (i.e., $\gamma_w = \gamma_i$) becomes $Q_b = \omega_0 / [2(\gamma_r + \gamma_i)] = Q_a / 2$, which is half of the lossless case. Hence, the maximum stored energy at the end of the transient regime is four times less than in the lossless case. In other words, the energy storage efficiency, which we define as the ratio of the cavity mode energy at any time instant to all energy spent on exciting the cavity up to that instant, will be significantly less than the case where the lossless cavity is not critically coupled to the radiation channel. After a transient period, all the input energy at every cycle of the excitation will be completely delivered into and dissipated in the cavity, making the steady state an undesirable regime to store energy in this cavity similar to the lossless cavity. In both scenarios, the excitation efficiency throughout the transient regime is quite low. Here we extend the notion of critical coupling to lossless cavities using non-monochromatic input signals oscillating at a complex frequency, inspired by the recently proposed concept of virtual absorption [17–19]. We demonstrate that, by tailoring the excitation field in time, the radiation channel can be critically coupled to a high-$Q$ resonant mode without the need for adding loss into the cavity to meet the critical coupling requirements. This *virtual critical coupling regime* leads to a perfect coupling of energy with unit coupling efficiency.

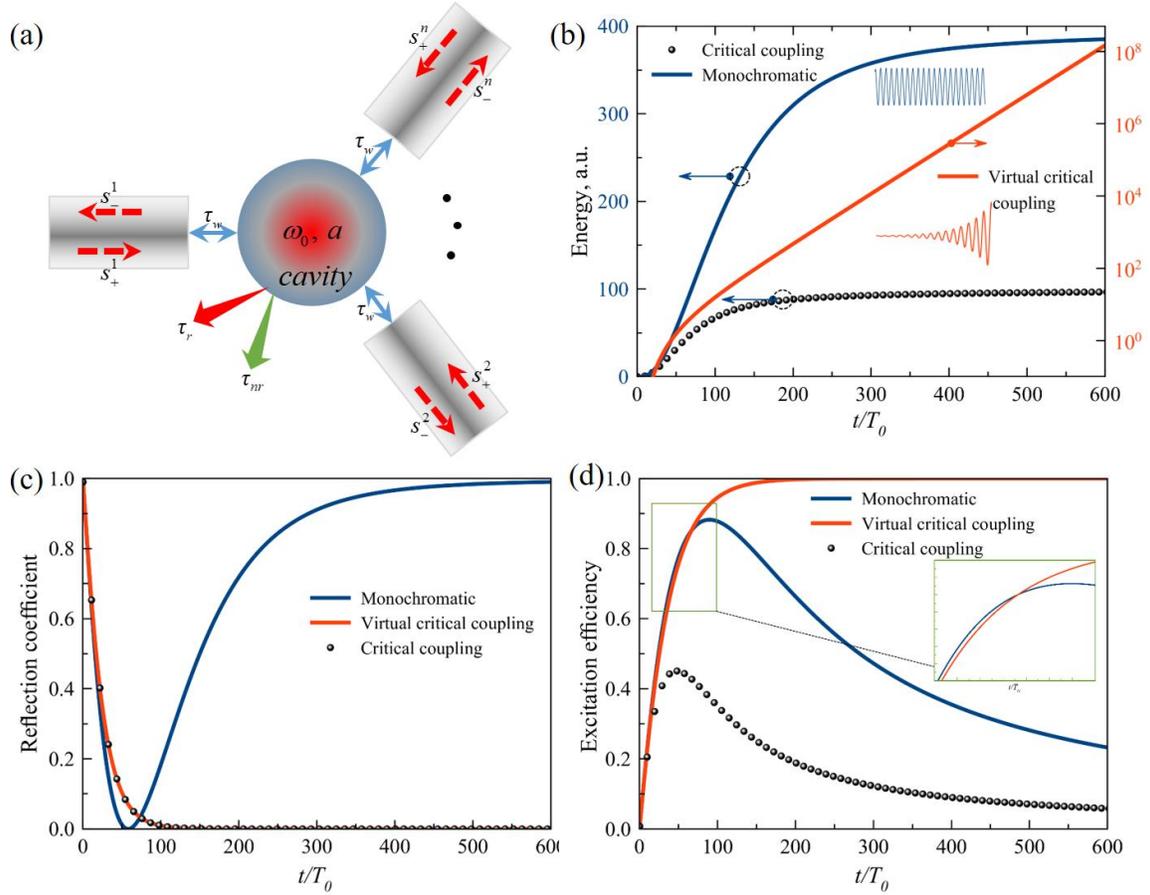

**Figure 1**. (a) Sketch for the virtual critical coupling analysis. (b) Energy in the cavity mode [note that the right axis in Fig. 1(b) is in logarithmic scale], (c) reflection coefficient, and (d) excitation

efficiency in the cases of monochromatic excitation [ $s_+^1(t) = t/(10T_0 + t)e^{-j\omega_0 t}$ ] of a lossless cavity with infinite intrinsic quality factor ( $Q_i \to \infty$ ) and the coupling quality factor $Q_w = 200$ (blue curves), monochromatic excitation of a critically coupled-cavity with $Q_i = Q_w = 200$ (black spheres), and virtual critical coupling [ $s_+^1(t) = 0.1t/(10T_0 + t)e^{-j\omega_0 t}e^{t/\tau_w}$ ] for a lossless cavity with infinite intrinsic quality factor ( $Q_i \to \infty$ ) and the coupling quality factor $Q_w = 200$ (red curves). The mode eigenfrequency is $\omega_0 = 2$, and the period is $T_0 = \pi$.

We start our analysis with a general description of critical coupling in the complex frequency plane. The scattering properties of a general resonant system, as sketched in Fig. 1(a), can be described rigorously by temporal coupled-mode theory (TCMT) [20,21]. We assume that the cavity has a single eigenmode at real frequency $\omega_0$. The mode amplitude $a$ is normalized such that $|a|^2$ corresponds to the stored energy in the cavity mode. In general, this resonant mode has nonradiative $\tau_{nr} = 1/\gamma_{nr}$ and radiative $\tau_r = 1/\gamma_r$ decay times (where the intrinsic quality factor and the decay rate are defined as $Q_i = \omega_0/2\gamma_i$ and $\gamma_i = \gamma_r + \gamma_{nr}$, respectively). The cavity is connected to $n$ channels with coupling constants $|\kappa\rangle = (\kappa_{w,1}, \kappa_{w,2}, ..., \kappa_{w,n})$, used for cavity excitation. The excited mode decays through these channels with a total decay rate $1/\tau_{w,tot} = \sum_{i=1}^{n} 1/\tau_{w,i}$, where $1/\tau_{w,i}$ is the decay rate to channel $i$. The coupling constants are related to the corresponding decay times as $\kappa_{w,i} = \sqrt{2/\tau_{w,i}} = \sqrt{2\gamma_{w,i}}$ [20,21]. As a result, the total decay rate ($1/\tau$) of the cavity is defined as $1/\tau = 1/\tau_{w,tot} + 1/\tau_{nr} + 1/\tau_r$. In contrast to common TCMT analysis, we distinguish the radiative decay to the excitation channels and to free space. We do so because the former is responsible for the cavity excitation, whereas the latter represents an additional loss channel defining the total $Q$. The evolution equation for the mode amplitude is [21,22]

$$\frac{da}{dt} = \left(j\omega_0 - \frac{1}{\tau}\right)a + \langle\kappa^*|s_+\rangle, \tag{1}$$

where $|s_+\rangle$ is the vector of input amplitudes (an $e^{j\omega t}$ time convention is assumed). It follows that, without excitation ($|s_+\rangle = 0$), the mode evolves in time as $a \sim e^{j\omega_0 t} e^{-t/\tau}$, i.e., it decays exponentially with complex frequency $\omega = \omega' + j\omega''$, where $\omega' = \omega_0$ and $\omega'' = 1/\tau$. Furthermore, the output amplitude vector $|s_-\rangle$ is related to the input vector and the resonance amplitude via [21,22]

$$|s_-\rangle = \hat{C}|s_+\rangle + a|\kappa\rangle, \qquad (2)$$

where $\hat{C}$ is the direct scattering matrix, which defines the direct pathway between input and output channels. In the following, we assume that there is no direct coupling between different channels and between a channel and free space without the cavity. Therefore, $\hat{C} = -\hat{I}$ where $\hat{I}$ is the unit matrix.

In the general case, the excitation signal $s_+^1 \sim e^{-j\omega t}$ is assumed to have a complex frequency $\omega = \omega' + j\omega''$, where the imaginary part $\omega''$ can be positive (exponentially growing in time) or negative (exponentially decaying). For a single port system, on which we focus in the following, this analysis gives $s_-^1 = -s_+^1 + a\kappa_w$ and $a = \kappa_w^* s_+^1 / [j(\omega' - \omega_0) + (1/\tau + \omega'')]$. Using these two expressions, we derive the reflection amplitude in the complex frequency plane as

$$r(\omega', \omega'') = \frac{(1/\tau_w - 1/\tau_{nr} - 1/\tau_r - \omega'') - j(\omega' - \omega_0)}{(1/\tau_w + 1/\tau_{nr} + 1/\tau_r + \omega'') + j(\omega' - \omega_0)}. \qquad (3)$$

Critical coupling for such a single-port system is achieved when $r(\omega', \omega'') = 0$, which can be satisfied if real and imaginary parts of the numerator equal zero, i.e., when $\omega' = \omega_0$ and $1/\tau_w = 1/\tau_{nr} + 1/\tau_r + \omega''$. For monochromatic excitations, i.e., signals with real frequency $\omega'' = 0$, this leads to the usual critical coupling condition $1/\tau_w = 1/\tau_{nr} + 1/\tau_r$, which is equivalent to $\gamma_w = \gamma_{nr} + \gamma_r$. In the case of no radiation loss, $1/\tau_r = 0$, this relation simplifies to the textbook critical coupling condition $1/\tau_w = 1/\tau_{nr}$ [22], in which the dissipative decay rate and the coupling rate of the excitation channel are equal.

From Eq. (1), it follows that the energy stored in the cavity at the end of the transient regime is $|a|^2 = \frac{4Q_w}{\omega_0} \frac{\gamma_w^2}{(\gamma+\omega'')^2} |s_+^1|^2$, where $\gamma = \gamma_w + \gamma_i$. In the case of monochromatic excitation ($\omega'' = 0$), the illuminated signal is $s_+^1 \sim s_0 e^{j\omega't}$, where the amplitude of excitation $s_0$ is a real constant. Therefore, the maximum energy that can be stored in the cavity at resonance equals $|a|^2 = \frac{4}{\omega_0} \frac{Q_w}{(1+Q_w/Q_i)^2} s_0^2$, where $Q_i = (Q_r^{-1} + Q_{nr}^{-1})^{-1}$ is the intrinsic quality factor. For cavities with large $Q_i$ (a cavity with small dissipation or radiation loss), the maximum stored energy at the end of the transient regime can be approximated as $|a|^2 \approx \frac{4Q_w}{\omega_0} s_0^2$ [23].

Figures 1(b)-1(d) exemplify this analysis for a lossless cavity with infinite intrinsic quality factor and a coupling quality factor of 200, for which we show the stored energy, the instantaneous reflection coefficient and the excitation efficiency as time evolves after the excitation ($t=0$). We start assuming a (real) frequency of excitation equal to the eigenfrequency of the cavity mode $\omega_0 = 2$ (blue solid lines in the Fig. 1). At early cycles of excitation, the structure is undercoupled and most of the illuminating energy is reflected at the input port. Only a tiny portion gets stored in the resonator, which is due to a strong mismatch between cavity and feeding channel, i.e., $\gamma_w \neq \gamma_i = 0$. However, as the excitation continues, the stored energy in the cavity grows (Fig. 1b) and, as a result, the outgoing wave stemming from the already-stored energy in the cavity becomes stronger and can cancel a portion of the instantaneous reflection of the input signal from the cavity input, resulting in a reduction of the total reflection coefficient (Fig. 1c). This process continues until an instant in time in which the structure is critically coupled (although the cavity is lossless) and the instantaneous reflection coefficient for monochromatic excitation actually vanishes [see Fig. 1(c)]. This point corresponds to the instant in time when the energy spillover due to the already-stored energy in the cavity is as strong as the direct reflection of the wave from the cavity input, and the two cancel out. This phenomenon indicates that there is an instant in time, during the transient excitation, in which the cavity is somehow critically coupled to the excitation despite being lossless, thanks to the fact that the mismatch at the input port is compensated by the spill-over driven by the energy already in the cavity. As the stored energy grows, however, this balance is lost and the structure becomes overcoupled; the reflection coefficient grows again until the cavity fills up and cannot store more energy [i.e., saturation point, see Fig. 1(b)], after which all the illuminating energy is reflected [see Fig. 1(c)].

In the time instant $\tau$ corresponding to this instantaneous critical coupling regime, the amplitude of the mode energy equals $|a(\tau)|^2 = |C|^2 \tau_w / 2$, as follows from Eq. (2), where $|C|$ is the direct scattering amplitude for the empty cavity. On the other hand, energy balance implies $\frac{d|a|^2}{dt} = |S^+|^2 - |S^-|^2$. Integrating this equation in time from 0 to $\tau$, and taking into account that in monochromatic excitation $|S^+|^2 = 1$, we get $|a|^2 = \tau\left(1 - \overline{|S^-|^2}\right)$, where the horizontal bar indicates average, which is approximately equal to $\overline{|S^-|^2} \approx |C|^2 / 2$ (assuming a linear variation of the reflection in this region). Substituting $|a|^2$ from the above expression, we get a simple approximate formula for $\tau \approx \frac{|C|^2}{2 - |C|^2} \tau_w$. For $|C|^2 = 1$, as in Fig. 1, this formula gives $\tau \approx \tau_w = 200$ or $\sim 64 T_0$. We note that the instant $\tau$ can be tailored by design, and may be moved down to very small values if $|C|^2 \sim 0$. The energy storage efficiency in this monochromatic excitation scheme, defined as the ratio of stored energy in the resonator at any time instant to the energy spent to excite it, $\eta = |a(t)|^2 \Big/ \int_0^t |s_+^1(t)|^2 dt$, is quite low, and can never reach unity [see Fig. 1(d), blue curve].

If we critically couple the cavity by adding loss, i.e., $Q_w = Q_i$ for $1/\tau_w = 1/\tau_{nr} + 1/\tau_r$ [24,25], the stored energy at the end of the transient regime reaches to $|a|^2 = \frac{Q_w}{\omega_0}$, four times lower than the lossless case, as seen in Fig. 1(b) (black dots). In this case, after the initial reflections at the start of the excitation, the cavity rapidly reaches very low reflections and converges to zero reflections in steady-state [Fig. 1(c)]. The storage efficiency in this scenario is lower than in the lossless cavity because the energy is continuously being dissipated.

Inspired by the instantaneous critical coupling regime highlighted at instant $\tau$ in the previous calculations, we now aim at considering complex frequency excitations, so that as the energy in the resonator grows, we increase proportionally also the excitation signals to maintain the system in the critically coupled regime, despite the absence of loss. In other words, we exploit the spill-over from the cavity to cancel the instantaneous reflections and continue pumping the cavity with maximum efficiency. In the complex frequency regime, we get a generalized form of critical coupling:

$$\omega' = \omega_0, \text{ and } \omega'' = \frac{1}{\tau_w} - \left(\frac{1}{\tau_{nr}} + \frac{1}{\tau_r}\right). \tag{4}$$

This complex frequency solution for critical coupling ensures zero reflection coefficient in (3). This scheme also allows to induce other regimes of interest. For instance, it is possible to realize an under coupled regime, i.e., $1/\tau_w > (1/\tau_{nr} + 1/\tau_r)$, by considering an excitation exponentially growing in time ($\omega'' > 0$), corresponding to a reflection zero in the upper complex frequency half-plane. Instead, we can realize an over-coupled resonator, i.e., $1/\tau_w < (1/\tau_{nr} + 1/\tau_r)$, by considering an exponentially decaying excitation ($\omega'' < 0$), corresponding to a reflection zero in the lower complex frequency half-plane. An extreme scenario is when the cavity is extremely under-coupled, i.e., a cavity with no dissipation or radiation loss. In this case, the solution of Eq. (4) reveals a fundamental difference between real and complex frequency excitation scenarios: in contrast to real frequency excitation (monochromatic excitation), the virtual critical coupling condition does not require the presence of dissipation or radiation loss to yield zero reflections from the cavity, or more generally move the reflection zeros within the complex frequency plane. Under the critical coupling condition, we avoid loss in the form of reflections from the cavity, and also avoid dissipation of the stored energy, simply because the cavity is lossless.

When Eq. (4) is satisfied, $\omega'' \neq 0$ and $|s_+^1|^2 \sim s_0^2 e^{2\omega'' t}$ (we assume that $|s_+^1|^2 = s_0^2$ at $t = 0$, where $s_0^2$ is small), based on the above analysis, the energy stored in the cavity reads $|a(t)|^2 = \frac{2\gamma_w}{(\gamma_w + \gamma_i + \omega'')^2} s_0^2 e^{2\omega'' t}$, which represents exponential growth of the stored energy in time as $|a(t)|^2 = |a(0)|^2 e^{2\omega'' t}$. The slope of this growth is defined by the factor $\frac{2\gamma_w}{(\gamma_w + \gamma_i + \omega'')^2}$, which has a maximum at $\gamma_w = \gamma_i + \omega''$, coinciding with the virtual critical coupling regime in Eq. (4).

The energy storage efficiency becomes $\eta = 4\gamma_w \omega'' / (\gamma_w + \gamma_i + \omega'')^2$, which in the virtual critical coupling regime simplifies to $\eta = 1 - \gamma_i/\gamma_w$. For a lossless cavity ($\gamma_i = 0$), it results in unitary energy storage efficiency $\eta = 1$ [see Fig. 1(d)], which implies that *all the incoming energy is stored in the cavity mode*. These results confirm the advantage of exciting the cavity at a complex frequency, in comparison to the previous scenarios. In principle, the cavity absorbs and stores all the incoming wave indefinitely with no reflection, no dissipation, and no saturation.

The red lines in Figs. 1(b)-1(d) show the result for the same lossless cavity considered above, excited now at this virtual critical coupling regime (4). At the early stages of excitation, the structure is undercoupled and the reflection coefficient is close to unity, however the reflected energy is almost zero simply because the exponentially growing excitation signal has a negligible amplitude in this region [see Figs. 1(b) and 1(c)]. As time goes by, the amplitude of the excitation signal grows, however at the same time, the reflection coefficient continues decreasing [see Fig. 1(c)], keeping the reflected wave negligible throughout the excitation. Therefore, although the energy storage efficiency is low at the initial stages of the excitation [see Fig. 1(d)], only a negligible amount of energy is lost in this transient regime. After the initial stage, the reflection coefficient continuously decreases until it reaches zero and stays zero simply because the incident wave is growing with the same rate as the scattering of the already-stored energy in the cavity [see Fig. 1(c)]. This time instant corresponds to the moment when the structure becomes critically coupled and stays in that state afterwards. The stored energy in the cavity grows exponentially [right axis in Fig. 1(b) is in logarithmic scale], following the excitation.

It is interesting to observe that, as can be seen from Figs. 1(c) and 1(d), there is a time interval at the early stages of the excitation where the reflection coefficient for the monochromatic excitation of the lossless cavity is smaller than the virtual critical coupling regime. During this time period, the efficiency of the monochromatic excitation of the lossless cavity is slightly larger than the virtual critical coupling case. The virtual critical excitation rapidly catches up after the instant $\tau$. It follows that an optimized excitation scheme may improve even further the response in the early stages of excitation, beyond the performance of the complex frequency excitation. For instance, we could first excite the cavity using a monochromatic excitation, until reaching the instant in which the reflection coefficient becomes zero, at $t = \tau$, and then start growing the excitation at the rate of growth of the stored energy, i.e., at the complex frequency of the reflection zero, to achieve an even higher storage efficiency. A comparison between this scenario and the original virtual critical coupling scenario is shown in Fig. 2. It is clear that, the compound scenario shows lower reflection coefficient [see Fig. 2(a)] and as a result the overall energy storage efficiency in this scenario is larger than the virtual critical coupling case [see Fig. 2(b)].

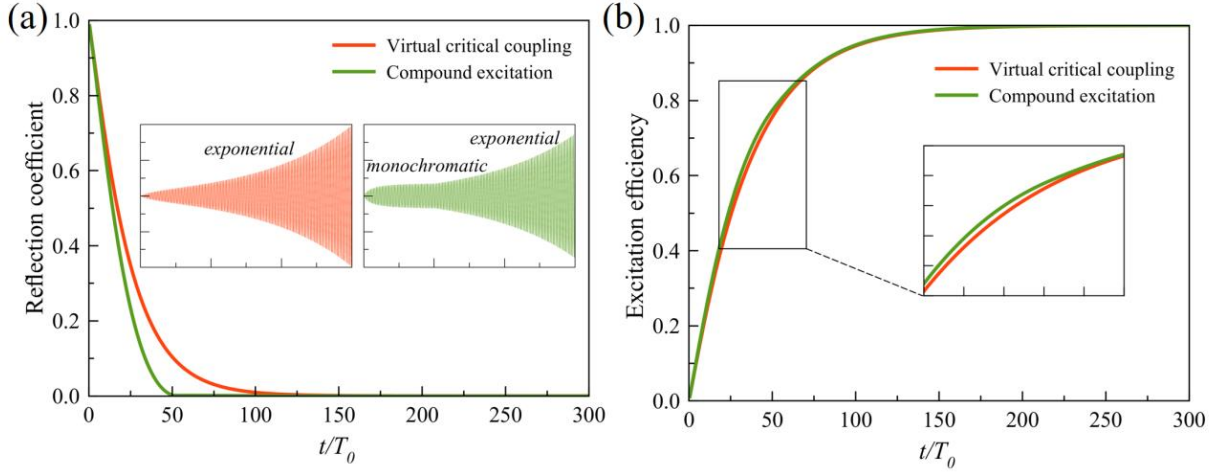

**Figure 2**. Comparison between compound excitation and virtual critical coupling scenarios. (a) Reflection coefficient and (b) excitation efficiency. Here, the compound excitation signal is:

$$s_+^1(t) = \theta(-t+50)\frac{t}{(5T_0+t)}e^{-j\omega_0 t} + \theta(t-50)e^{-j\omega_0 t}e^{(t-50)/\tau_w}.$$

Figure 3(a) shows a schematic of the mechanism of complex frequency excitation of a one-port cavity composed of a dielectric layer sandwiched between a perfect electric conductor and a highly reflecting mirror with no dissipation loss. A few observations can be made from the results shown in Fig. 1 for the virtual critical coupling in such a system: at the early stages of excitation, the amplitude of the reflection coefficient (i.e., $R = |-1 + a\kappa_w/s_+^1|^2$) is close to unity since there is no stored energy in the cavity yet. After a transient period, as it is shown in Fig. 3(b), the stored energy grows strong enough so that its leakage from the cavity cancels the backscattering of the incident wave from the cavity (i.e., $|a\kappa_w/s_+^1|^2 \to 1$). After this transient regime, since the illuminated signal is hitting the right complex zero of the system, the stored energy and the incoming wave grow at the same pace, ensuring that this total destructive interference continues indefinitely.

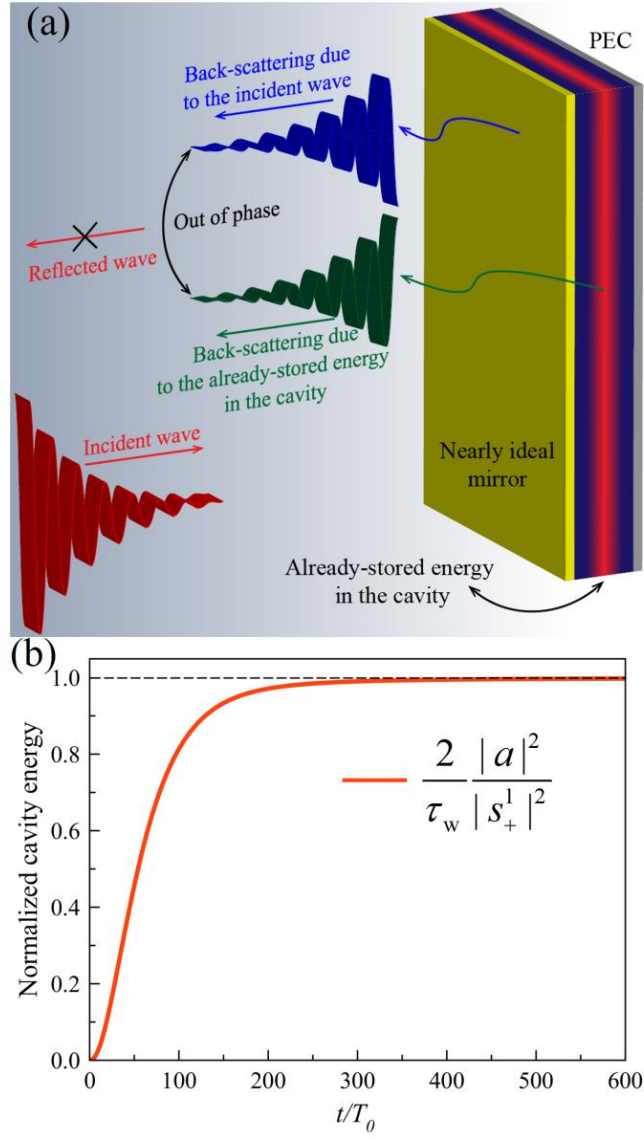

**Figure 3**. (a) Schematic of a system designed based on the virtual critical coupling approach. (b) Illustratively explanation of the physical mechanism behind the energy storage in the virtual critical coupling regime [ $s_+^1(t) \sim e^{-j\omega_0 t} e^{t/\tau_w}$ ].

Although we have considered so far a single-port cavity, relevant for perfect absorbers [24,26], antenna systems [27,28], and light coupling to ultrathin and 2D materials [29–31], a similar analysis can be generalized to multiport cavities. The analysis of Eqs. (1) and (2) in the complex frequency plane shows that the concept of virtual critical coupling can be generalized to systems with multiple excitation channels. In such cases, the total coupling rate can be written as $1/\tau_{w,tot} = \sum_{i=1}^{n} 1/\tau_{w,i}$, and the cavity mode should be excited by a set of complex waves coherently from all channels matching the eigenmode corresponding to the reflection zero of choice. We note that in this multi-port scenario, the critical coupling regime corresponds to zero reflection in all channels simultaneously [19,25].

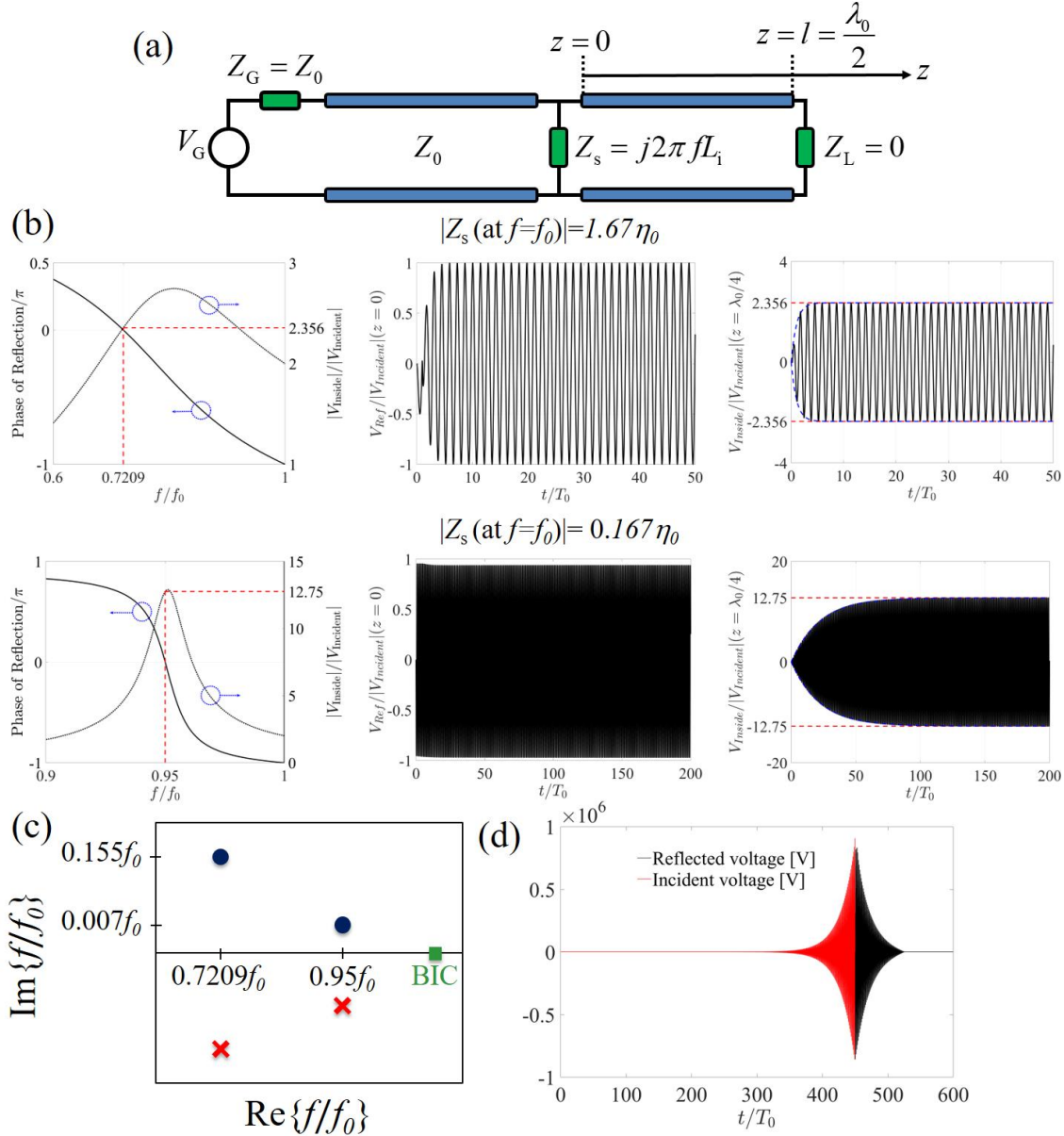

**Figure 4.** (a) Transmission line model for a high-$Q$ cavity closed by the inductor $L_i$. (b) Phase of reflection, the amplitude of reflection, and amplitude of voltage in the middle of the cavity for different $Z_s$. These results show that the larger the $Q$-factor, the longer the time it needs to excite the cavity. Moreover, there is a maximal amount of energy that can be stored in the cavity. (c)

Location of the complex zeros and poles for different values of $Z_s$. Reducing $L_i$ leads to the moving of the poles and zeros of the reflection coefficient towards the real frequency axis, giving rise to larger $Q$ of the resonant mode. At very small values of $L_i$ the pole and zero eventually coalesce resulting in BIC state formation. (d) Time-domain input and output signals in the virtual critical coupling regime for the case $|Z_s(\text{at } f = f_0)| = 0.167\eta_0$.

Next, we investigate the applicability of the virtual critical coupling regime in a practical waveguiding system. Let us consider a transmission-line cavity, as shown in Fig. 4(a). The cavity is shorted at one end and is closed using a reactive load $Z_s$ at its other end. We note that such a system is widespread in RF and microwave technologies, and similar RLC analysis can be applied for the investigation of optical cavities [22]. Firstly, we study the temporal dynamics of the cavity under harmonic excitation. In the Laplace domain, the reflection from such a cavity reads

$$V_{\text{Ref}}(0, s) = \frac{\omega_0}{s^2 + \omega_0^2} \left[ -\frac{Z_0}{2Z_s + Z_0} - \left(\frac{2Z_s}{2Z_s + Z_0}\right)^2 \sum_{n=0}^{\infty} \left(\frac{Z_0}{2Z_s + Z_0}\right)^n e^{-\frac{2(n+1)l}{c}s} \right], \quad (5)$$

where, $\omega_0$ and $Z_0$ are the excitation frequency and characteristic impedance of the transmission line, respectively. By taking the inverse Laplace transform, we can get the time-domain reflections from the cavity as we excite it with signals with real frequencies. In steady-state, the reflection coefficient can be calculated as

$$r = \frac{-[\omega L_i + \eta_0 \tan(\beta l)] + j\omega L_i \tan(\beta l)}{[\omega L_i + \eta_0 \tan(\beta l)] + j\omega L_i \tan(\beta l)} \quad (6)$$

The amplitude of the total voltage in the middle of the transmission line cavity at steady-state reads $V_{\text{Inside}}(z = l/2) = \frac{1+r}{2}\sec\left[\frac{\beta l}{2}\right]$. Figure 4(b) shows two different scenarios where the lossless cavity is closed with different reactive loads. Naturally, a smaller reactive load, closer to a short circuit, provides a higher quality factor and a longer time to reach the steady-state, allowing more energy storage in the cavity. Figure 4(b) shows the frequency response, the time evolution of the reflected waves as we start exciting the cavity, and the corresponding time evolution of the voltage in the middle of the transmission line for cavities with different $Q$ factors and linewidths. During each cycle in the transient, some energy gets stored in the cavity and the rest gets reflected. The reflection from the cavity grows as more energy is stored in the cavity, and, as the system reaches steady-state, the cavity cannot store more energy and the system fully reflects the incoming signal.

Next, we utilize the above analysis and apply the virtual critical coupling concept for the excitation of this resonator. Figure 4(c) shows the zeros and poles of the reflection coefficient of the system in the complex frequency plane for two different scenarios considered in Fig. 4(b). As the inductive layer gets closer to an ideal perfect mirror, the complex zero of the system gets closer to the real frequency axis, giving rise to a bound state in the continuum (BIC) [15,16,19,32–34]. Note that this specific scenario is consistent with the recently proposed epsilon-near-zero (ENZ)-based BICs arising at the plasma resonance when the ENZ becomes nontransparent and the corresponding mode becomes bounded [33] or its implementation with resonant metasurfaces [32]. In the limit of a very small inductive load and very high $Q$-factor, one can excite the system with a semi-harmonic excitation (i.e., a harmonic excitation with the very low growth rate). This can be of interest in practical applications where it might be challenging to continue a growing signal for a long period of time. Figure 4(d) shows the scenario in which our input signal engages the first complex zero of the second case in Fig. 4(b). Similar to what was predicted in the theoretical analysis, in contrast to the monochromatic excitation, while the input signal excites the cavity there is no reflection from the system and the cavity can be charged with unitary efficiency.

*Conclusions*. – In this paper, we have generalized the concept of critical coupling to the case of complex frequency excitation in order to improve the efficiency of the excitation of high-$Q$ resonators. Based on this approach, we can temporally shape the excitation to make sure that a lossless high-Q cavity fully absorbs all impinging energy without any reflection. The absorbed energy will be stored in the cavity without getting dissipated and can be released at will. Remarkably, we have shown that this approach of virtual critical coupling is the only possible way to achieve unit excitation efficiency in open systems. We have demonstrated the applicability of the approach in the transmission line system, and envision similar pathways for optical and nanophotonic systems. Our work may lead to the efficient excitation of high-Q resonators and the generation of efficient memories and energy storage.

This work was supported by the Air Force Office of Scientific Research and the Simons Foundation.